\documentstyle[11pt,epsfig]{article}

\newcommand \be{\begin{equation}}
\newcommand \bea{\begin{eqnarray}}
\newcommand \ee{\end{equation}}
\newcommand \eea{\end{eqnarray}}

[1999/03/29 PSNFSS v.7.2
Times font as default roman
: S Rahtz]

\begin{document}

\title{Fearless versus Fearful Speculative Financial Bubbles}

\author{J. V. Andersen $^{1,2}$ and D Sornette$^{1,3}$\\
$^1$ Laboratoire de Physique de la Mati\`{e}re Condens\'{e}e\\ CNRS UMR6622 and
Universit\'{e} de Nice-Sophia Antipolis\\ B.P. 71, Parc
Valrose, 06108 Nice Cedex 2, France \\
$^2$ U. F. R. de Sciences \'Economiques, Gestion, Math\'ematiques et
Informatique, \\ CNRS UMR7536 and Universit\'e Paris X-Nanterre,
92001 Nanterre Cedex, France \\
$^3$ Institute of Geophysics and
Planetary Physics and Department of Earth and Space Science\\ 
University of California, Los Angeles, California 90095\\
e-mails: vitting@unice.fr and sornette@unice.fr\\}

\date{\today}
\maketitle

\vskip 2cm
{\bf Abstract:} 
Using a recently introduced rational expectation model of bubbles, based
on the interplay between stochasticity and
positive feedbacks of prices on returns and volatility,
we develop a new methodology to test how this model
classifies 9 time series that have been
previously considered as bubbles ending in crashes.
The model predicts the existence of two anomalous behaviors 
occurring simultaneously: (i) super-exponential price growth and (ii)
volatility growth, that we refer to as the ``fearful singular bubble''
regime. Out of the 9 time series, we find that 5 pass our tests and
can be characterized as ``fearful singular bubbles.'' The 4 other cases
are the information technology Nasdaq bubble and three bubbles of the Hang Seng
index ending in crashes in 1987, 1994 and 1997. According to our analysis, these
four bubbles have developed with essentially
no significant increase of their volatility. This paper thus
proposes that speculative bubbles ending in crashes form two groups hitherto
unrecognized, namely
those accompanied by increasing volatility (reflecting increasing risk perception)
and those without change of volatility (reflecting an absence of risk perception).


\newpage

\section{Introduction on the detectability of bubbles}

Looking back at the financial bubble of 
the late 1990's, questions are now increasingly
raised as to why people did not realise in time the existence of 
a bubble and why nobody acted
early enough to prevent the maniacal increase in stock prices and the investment frenzy
in technological businesses that happened during the bubble. 
Contradictory statements abound. 
Greenspan's 1996 catchphrase, ``irrational exhuberance,''
interpreted to refer to an overvalued U.S. stock market is often 
remembered as a premonitory insight \cite{Green1}. In a
private meeting, Greenspan told Lawrence Lindsey (former
Fed Governor) that he saw ``all sorts of parallels to the late 1920s''
(quote from Investor's Business Daily, 8/24/1999).
Similarly, ``The Greenspan model unveiled by the Federal Reserve chairman in 1997,
calculates that the S\&P is a whopping 67\% above its appropriate level.''
(quote of Andrew Bary, Barron's, 1/24/2000).
Robert Rubin, the chairman of the executive committee of Citigroup
Inc., the biggest US financial services company, said ``the world faces
serious and continuing danger of another economic crisis like Asia's
in 1997 and 1998. Threats to wider world trade, the growing gap between
rich and poor nations and unwise investments based on the idea that new
technology will ensure prosperity all could set off such a crisis...
The idea that new technologies can erase business cycles is a myth,''
Rubin said. ``New technologies are of profound importance, but they are
not the first new technologies of significance. Autos,
electricity, railroads and medicine led earlier productivity booms, 
and none of them, separately or together, produced one-way
prosperity.'' (Bloomberg News, quoted by William Fleckenstein, The
Contrarian, 2/7/2000).
Michel Camdessus, outgoing head of the International Monetary Fund, said
``Overvalued US share prices and low US savings rates are two risks to
an otherwise rosy world economic picture... you have this extreme risk of
dramatically insufficient domestic savings translating into a current
account deficit of a very great size... And there is indeed a risk of an
overvalued financial market -- by any conventional standards -- which
create a situation of vulnerability of which we must be mindful.''
(Michel Camdessus to Janet Gittsman, Reuters, 2/8/2000). All these
statements were issued before the crash on the Nasdaq in March-April 2000 and
before the start of the descent of all major US markets (as well as most
western markets).

In contrast, consider the following statements by A. Greenspan, issued
after the burst of the last two notworthy bubbles.
``We did not foresee such a breakdown in Asia. I suspect that the very
nature of the process may make it virtually impossible to anticipate. It
is like water pressing against a dam. Everything appears normal until a
crack brings a deluge'' (Fall 1998) \cite{Green2}.
``The struggle to understand developments in the economy and 
financial markets since the mid-1990s has been particularly
challenging for monetary policymakers. We were confronted with forces
that none of us had personally experienced. Aside from the then recent
experience of Japan, only history books and musty archives gave us clues
to the appropriate stance for policy. We at the Federal Reserve
considered a number of issues related to asset bubbles--that is, surges
in prices of assets to unsustainable levels. As events evolved, we
recognized that, despite our suspicions, it was very difficult to
definitively identify a bubble until after the fact--that is, when its
bursting confirmed its existence. Moreover, it was far from obvious that
bubbles, even if identified early, could be preempted short of the
central bank inducing a substantial contraction in economic
activity--the very outcome we would be seeking to avoid'' \cite{Green3}.
In the same conference in August 2002, 
A. Greenspan added ``The notion that a well-timed incremental
timing could have been calibrated to prevent the late 1990's bubble is 
surely an illusion.'' However if money supply growth had been kept in check,  
interest rates would have risen, and people would have been more careful 
with their investment choices, which would have had a dampening effect on 
the frantic rising stock markets. But, before one can use money supply as a financial 
instrument, it requires a knowledge that one is actually in the 
middle of a bubble phase. In hindsight, this could sound like an easy task, but 
the above experiences suggest that even the most skilled professionals have a very 
hard time.

The situation on the detection of bubbles
is perhaps even murkier in the academic literature. Empirical
research has largely concentrated on testing for 
``explosive'' exponential trends in the time
series of asset prices and foreign exchange rates \cite{Evans,Woo}, with
however limited success. The first reason lies
in the absence of a general definition, as bubbles are model specific and generally
defined from a rather restrictive framework.  The concept of a fundamental
price reference does not necessarily exist, nor is it necessarily unique. 
A major problem is that apparent evidence for
bubbles can be reinterpreted in terms of market fundamentals that are
unobserved by the researcher. Another suggestion is that, if
stock prices are not more explosive than dividends, then it can be
concluded that bubbles are not present, since bubbles are by definition
the explosive component to stock prices not explained by dividends.  However, 
periodically collapsing bubbles are not detectable by using
standard tests to determine whether stock prices are more explosive or
less stationary than dividends \cite{Evans}. Some recent attempts use
for instance Markov-switching models to distinguish a high-return stable state
from a low-return volatile state \cite{Maheu} or
variance bounds tests, bubble
specification tests, and cointegration tests on equity prices and dividends 
\cite{Brooks}, or measures of deviations of future from expected returns 
\cite{Siegel}.

Here, we present a different approach based on the idea that a bubble
is more than just an exponential growth, it is a super-exponential growth.
We thus use a recently introduced rational expectation model of bubbles,
capturing the interplay between stochasticity and nonlinear positive feedback
of prices onto future returns and volatility. We use this model to
develop a methodology for the detection of bubbles. The next section recalls the formulation of the
model and formulates the hypothesis and the associated null hypothesis.
Section 3 describes the methodology of the test and its implementation.
Section 4 describes the results obtained on 9 time series that have been
previously argued to be speculative bubbles ending in crashes. Section
summarizes our results and concludes.

\section{Formulation of the nonlinear rational expectation bubble model}

One can summarize the formulations of previous models of bubbles $B(t)$ as
\be
{d B(t) \over B(t)} = \mu dt + \sigma dW_t - \kappa  dj~,  \label{nfkaaak}
\ee
where $B(t)$ is the price above the fundamental price,
$\mu$ is the abnormal return rate above the fundamental return,
$\sigma$ is the volatility of the bubble and
the jump term $dj$ describes a correction or a crash
that may occur with amplitude $\kappa$.
As we already mentioned, the problem is that (\ref{nfkaaak}) is
of the same form as the standard geometrical Brownian motion often used to model the
fundamental price. The qualification of a bubble in this context boils
down to measuring an abnormal exponential rate $\mu >0$ of growth, a
rather uncontrolled procedure since
the fundamental growth rate acting as a reference is itself not known. 

In Ref.~\cite{firstpaper}, we have proposed to generalize (\ref{nfkaaak})
by allowing $\mu(B(t))$ and $\sigma(B(t))$ to become
nonlinearly dependent on the instantaneous realization of the price.
Specifically,
\bea
\mu(B) B &=& {m \over 2B} [B \sigma(B)]^2 + \mu_0 [B(t)/B_0]^m ~,  \label{buyaauqka}   \\
\sigma(B) B &=&  \sigma_0 [B(t)/B_0]^m~,  \label{non_lin_rel}
\eea
where $B_0$, $\mu_0$, $m>0$ and $\sigma_0$ are four parameters of the model, setting respectively
a reference scale, an effective drift, the strength of the nonlinearity
and the magnitude of stochastic component which sets the scale
of the volatility. 
The first term in the r.h.s. (\ref{buyaauqka}) is added as 
a convenient device to simplify the Ito calculation of these stochastic differential
equations. In  Ref.~\cite{firstpaper}, we discussed several mechanisms
that justify the positive feedback 
of $\mu(B(t))$ and $\sigma(B(t))$ on stock prices. Herding is perhaps the 
most obvious mechanism that leads to a positive nonlinear feedback, since 
the mere fact that the price of a stock rises/falls attracts more
buyers/sellers respectively. Reputation herding, information cascades, the 
wealth effect and hedging strategies are other sources of nonlinear
positive feedbacks of the type described by (\ref{buyaauqka},\ref{non_lin_rel})
(see \cite{firstpaper,bookcrash} and references therein).

The model (\ref{nfkaaak}) with (\ref{buyaauqka},\ref{non_lin_rel}) is a
rational expectation model of bubbles \cite{Blanchard,Blanwat}.
The feature that distinguishes it from ordinary bubble models 
is that the solutions may be at some times growing ``super-exponentially,''
i.e., with a growth rate growing itself with time. 
This is fundamentally different from
previous bubble models based on exponential growth (with constant
average growth rate). This suggests to qualify bubbles by
testing for the existence of nonlinear super-exponential growth.
Similar ideas have been previously developed by combining the
rational expectation theory of bubbles together with models of
collective agent behavior \cite{ledoit1,Ledoit2}
(see also \cite{bookcrash,SorJoh,SorPR} and references therein).

The model  (\ref{nfkaaak}) with (\ref{buyaauqka},\ref{non_lin_rel})
can be reformulated in the Stratonovich interpretation
given by the expression:
\be
{d B \over dt} = (a \mu_0 + b \eta)~B^m -\kappa B dj~,  \label{jfja}
\ee
where $a$ and $b$ are two positive constants and $\eta$
is a delta-correlated Gaussian white noise. The form (\ref{jfja})
examplifies the fundamental ingredient of our theory
based on the interplay between nonlinearity and multiplicative noise. 

The solution of (\ref{nfkaaak}) with (\ref{buyaauqka}) and (\ref{non_lin_rel})
is derived in \cite{firstpaper} and reads
\be
B(t) = \alpha^{\alpha}~{1 \over \left(\mu_0[t_c - t] - 
{\sigma_0 \over B_0^{m}}~W(t)\right)^{\alpha}}~,  ~~~~{\rm where}~\alpha\equiv {1 \over m-1}
\label{sb}
\ee
with $t_c= y_0/(m-1)\mu_0$ is a constant determined by the initial condition
with $y_0=1/[B(t=0)]^{m-1}$. Expression (\ref{sb}) is correct as long as 
a crash $dj=1$ has not occurred, which may happen at any instant according
to the crash hazard rate 
\be
h(t) = {\mu(B(t)) \over \langle \kappa \rangle}~,  \label{bvfjuaj}
\ee
determined from the no-arbitrage condition. Here, $\langle \kappa \rangle$ is
the average amplitude calculated over some pre-determined distribution of $\kappa$.
When a crash occurs, $B(t)$ is transformed into $B(t)(1-\kappa)$ which provides
a novel initial condition which is plugged in (\ref{nfkaaak}) with (\ref{buyaauqka}) 
and (\ref{non_lin_rel}).

In the deterministic case $\sigma_0=0$, 
(\ref{sb}) reduces to $B(t) \propto 1/[t_c-t]^{1 \over m-1}$, i.e., the
bubble follows a hyperbolic growth path which would diverge in finite time, if not
checked by crashes according to (\ref{bvfjuaj}). Note that
this hyperbolic growth is the signature of the
positive feedback characterized by $m>1$ of the price $B(t)$ on the return rate $\mu$.
Reintroducing the stochastic component $\sigma_0 \neq 0$, we see
from (\ref{sb}) that a ghost finite-time singularity still exists but
its visit is controlled by the first passage of a biased random walk at the 
position $\mu_0 t_c$ such that the denominator
$\mu_0[t_c - t] - {\sigma_0 \over B_0^{m}}~W(t)$ vanishes. Due to the
crash hazard rate which grows even faster than does the bubble price, 
this ghost-like singularity is never attained. For extensions on 
the concept of noisy finite-time singularities similar to (\ref{sb}), see
Ref.~\cite{Fogedbty1,Fogedbty2}.

Taking the limit $1/\alpha \rightarrow 0$ ($m \to 1$) in (\ref{sb}) recovers
the standard Black-Scholes solution
\be
B_{\rm BS}(t) = \exp{\left(\mu_0 t +  \sigma_0 W(t)\right)}~.
\label{bs}
\ee
This is expected from
(\ref{nfkaaak}) with (\ref{buyaauqka},\ref{non_lin_rel}) for $m=1$
which has the standard differential form of a geometric random walk.

\section{Testing for nonlinear bubbles}

\subsection{Formulation of the bubble detection test \label{what}}

Using the two expressions (\ref{sb}) and (\ref{bs}), we propose 
the following testing procedure in order to qualify a given time series as a bubble.
We want a test that distinguishes between a standard biased geometric
Brownian motion (\ref{bs}), called from now on the
Black-Scholes (BS) model taken as the benchmark for ``normal'' times,
and the stochastic hyperbolic or ``singular bubble'' (SB)  regime (\ref{sb}).
Since the BS solution (\ref{bs}) is embedded in the more general SB 
expression (\ref{sb}) as the special case $1/\alpha \rightarrow 0$, 
testing for the presence of a bubble amounts to test whether 
$1/\alpha$ is different from $0$ with a sufficient high confidence.
A first step is to resort to the general method of testing for embedded hypotheses
and use the Likelihood Ratio test (Wilks' chi-square statistics) \cite{Wilks}.

In our empirical tests, we shall focus on time series which have been argued
in the past to be bubbles because prices did appreciate by large margins before
crashing precipitously over a short time interval. They are shown 
in figures 1-9. The table below each figure gives the studied time period,
the SB and BS acceptance percentages defined below, and the values of the fit
parameters for the plots shown in the figures. The crash happens right after the final date 
for each time series. By focusing on the part of
the price trajectory before the crash, we do not need to worry about the
crash jump process and do not need to address the impact of the crash hazard rate. 

We assume that the observed price $P(t)$ evolves according to
\be
\ln P(t) = rt + \ln[F] + \ln[B(t)]~.   \label{P}
\ee
Expression (\ref{P}) accounts for the presence of a fundamental component
$F$ and for the existence of an exogenous growth rate $r$ of the economy.
This formulation (\ref{P}) is different from our previous model
in which $F$ was taken additive to $B$ \cite{firstpaper}. We propose that
(\ref{P}) is more appropriate since $\ln[B(t)]$ follows a geometric Brownian
motion according to the null BS hypothesis and
$\ln (F)$ thereby gives the constant level of the random walk proportional to $W(t)$.

The SB hypothesis is
defined by the model (\ref{P}) with (\ref{sb}) which has 
six parameters $(\alpha, F, r ,t_c, \mu_0, \sigma_0)$.
The BS null hypothesis is specified by the model (\ref{P}) with (\ref{bs}) 
and has three parameters $(F, r+\mu_0, \sigma_0)$. Note that the two
parameters $r$ and $\mu_0$ which play distinct roles in the SB model can be
fused into a single parameter $r+\mu_0$ for the BS model as seen
from equations (\ref{bs},\ref{P}), making it effectively
a three-parameter model.

What are we really testing when comparing the SB with the BS model?
As can be seen from expression (\ref{sb}) and from (\ref{buyaauqka}),
the first characteristics of a SB is its potential for 
faster-than-exponential accelerated growth, 
as we already stressed. There is another important 
undissociated characteristics, as shown by expression (\ref{non_lin_rel}), that
the volatility of a SB also tends to increase and to accelerate as the bubble
increases. Thus, our tests for the detection of SB are 
probing the existence of simultaneous stochastic accelerations of the price and of the
volatility. Our model has thus a behavior superficially in contradiction with the so-called
leverage effect, in which volatility seems
to rise when a stock price drops, and fall when the stock goes up
(see the early reference \cite{Black} and \cite {Figlewski,bouchaud} and references therein).
The leverage effect is usually identified 
by a careful average over many different time series, usually over 
many decades of data and is thus believed 
to characterize the stationary behavior of prices.
The leverage effect does not exclude 
the possible existence of isolated periods of positive correlation 
between past returns and future volatility.
Indeed, while the leverage effect describes a short-lived correlation between past
returns and future volatility, our emphasis is the
cumulative trends and
accelerations both in prices and volatility occurring on the longer time scales.
Our SB model thus emphasizes the possibility for two anomalous behaviors 
occurring simultaneously: (i) super-exponential price growth and (ii)
accompanying volatility growth. We shall refer to this behavior as
the ``fearful SB,'' which is characterized by an increase of risk aversion
as the bubble develops.
It can thus be expected that some market
phases will exhibit only one of the two, for instance only an 
acceleration of price without increasing volatility or even with 
decreasing volatility. 
In this case, our model may reject the SB hypothesis.
The coexistence of the two characteristics (i) and (ii)
can be summarized pictorially by the sentence: ``Bull
markets climb a wall of worry, while Bear markets decline on a slope of
hope'' (citation attributed to Robert Pretcher). We are going to show
that this is not always the case.

\subsection{Implementation \label{impl}}

We describe the procedure for the SB case
(the BS case follows a similar procedure). We first use a Maximum
Likelihood procedure to estimate the parameters $(\mu_0,\sigma_0)$. From the initial 
value of the financial data, $t_c$ can in turn be 
determined. The remaining three parameters $(\alpha, F, r)$ 
are then determined by a goodness-of-fit to the financial data. 
This improves significantly on the inversion method used in our
previous work \cite{firstpaper}, based a Kolmogorov-Smirnov test of the
Gaussianity of the increments $dW$ obtained by inverting the analog of
(\ref{P}) with an additive fundamental contribution $F$ to the price $P$. 
Our previous calibration of the model used only the
information on the one-point statistics (distribution) of increments.
We have also tried to extend this methodology by developing tests probing
the residual correlation structure of the inverted increments $dW$ in the spirit of
\cite{Gour} but found this approach worse that our first method based
on ensuring the Gaussianity of the distribution of the increments $dW$.
Our present procedure performs even better than our first one
by taking into account simultaneously the
information in all multi-point statistics of the increments.

Define the variable $X(t)$ as:
\be
X(t) =\left[ P(t) e^{-rt}/ F  \right]^{-{1 \over \alpha}} ~.   \label{X}
\ee
If the observable price $P(t)$ was completely specified by our 
positive feedback model (\ref{sb}), the increments $d X(t)$ would 
be distributed according to a Gaussian distribution with variance 
$-{\sigma_0 \over \alpha} dt$ and mean ${\mu_0 \over \alpha} dt$. The likelihood 
of $N$ increments is known explicitely from the model
\be
L= \prod_{i=1}^N~p_L(dX_i)~,
\label{L}
\ee
where 
\be
p_L(dX) ={1 \over \sqrt{2 \pi (\sigma/\alpha)^2}} \exp
\left( {-\left( dX-(\mu_0/\alpha) dt \right)^2 \over
2(\sigma_0/\alpha)^2 dt }\right)~.   \label{P_LX}
\ee
The maximum likelihood value of $\mu_0$ is thus
\be
\mu_0/\alpha  = {1 \over N dt} ~\sum_1^N dX(i)~,
\label{mu}
\ee
where $N$ is the length of the time series of increments and $T=N dt$ its total
duration. Given fixed values of the 
parameters $(\alpha, F, r)$, $\mu_0$ can then be determined by (\ref{X}) 
and (\ref{mu}).

Similarly, let us define a variable $Z$ as
\be
Z(t) = X(t) -\mu_0 \left(t_c - t\right )~.   \label{Z}
\ee
The probability density distribution of the increment $dZ$ is 
\be
p_L(dZ) ={1 \over \sqrt{2 \pi (\sigma_0/\alpha)^2}} \exp
\left( {-\left( dZ \right)^2 \over
2\left(\sigma_0/\alpha \right)^2 dt} \right)~.  \label{P_LZ}
\ee
Constructing the likelihood as $\prod_{i=1}^N~p_L(dZ_i)$, its maximization
with respect to $\sigma$ gives
\be
(\sigma_0/\alpha)^2  = {1 \over N dt }~\sum_1^N [dZ(i)]^2  ~.
\label{sigma}
\ee
Given a fixed parameter set $(\alpha, F, r)$, one first determines
$\mu$ from (\ref{mu}). 
$\sigma$ can then be determined by (\ref{Z}) and (\ref{sigma}).

Using the initial value of the observable price $P(t=0)$, $t_c$ is then
determined by:
\be
t_c  =  {\alpha \over \mu_0}~[F/P(t=0)]^{1/\alpha}
\label{Tc}
\ee

The maximum likelihood method applied to the three other parameters
$(\alpha, F, r)$ leads to a set of nonlinear equations that have
no analytic solutions and are not conveniently dealt with numerically.
We have resorted to a non-standard method which consists literally to
perform an OLS (ordinary least-square) fit of the price
time series by the model (\ref{P}) with (\ref{sb}). Specifically, we have to find 
the minimum in the three dimensional parameter space of 
$(F,r,\alpha)$ of the expression: 
\be
Q_{\rm sb} \equiv \sum_{i=1}^N \left [P_{\rm data}(i) - P(i) \right ]^2/N ~.
\label{Q}
\ee
In practice, this search is performed by scanning of the 
parameters $(\alpha, r)$ and then searching for the minimum of 
$Q_{\rm sb}$ in (\ref{Q}) as a function of $F$. For each fix value of 
$(F,r,\alpha)$, $\mu_0$ is first calculated from (\ref{mu}) 
after which $\sigma_0$ can be determined from (\ref{sigma}) 
and $t_c$ from (\ref{Tc}). Subsequently, $Q_{\rm sb}$ can then 
be calculated from (\ref{Q}). 

The non-standard nature of this OLS fit stems from the fact that the model
$P(i)$ depends on the realization of the stochastic component $W(t)$.
For each fixed values of the parameters $(\mu_0, \sigma_0, t_c, F, r)$, we
generated $n$ different configurations of $W$.
$Q_{\rm sb}$ was then determined as the minimum over the $n$ 
realizations. The choice $n=5$ is a compromise
between a sufficient exploration of the space of realizations
and numerical feasibility. Increasing the number $n$ of stochastic
realizations too much is hindered by the slow convergence in the 
three-dimensional search space $(\alpha, F, r)$ due precisely to
the existence of the stochastic component of the model.

By applying the same procedure to the BS model (\ref{P}) with (\ref{bs}),
we obtain $Q_{\rm bs}$, defined as the minimum of the expression 
equivalent to (\ref{Q}) for the BS model over the $n$ stochastic realizations.
In principle, we can then use Wilks' theorem \cite{Wilks} 
on nested hypotheses and calculate the log-likelihood-ratio
\be
T  \equiv  -2 N (\log{\sqrt{Q_{\rm sb}}} - \log{\sqrt{Q_{\rm bs}}})~.
\label{T}
\ee
The log-likelihood-ratio allows us to test the SB model against the BS model. 
A large value of $T$ means that the SB model gives a strong improvement
over the BS model in fitting the data. To assess the statistical significance
of a given value $T$, we can use the standard result that
$T$ is asymptotically a $\chi^2$ variable with three degrees of freedom, because 
the BS model has three less parameters than the SB model.
The Wilks stastitics then gives the probability $p(T)$ that
the observed value of $T$ can be surpassed by chance alone, given
the confidence level $1-p(T)$ at which the SB model is prefered over
the null BS hypothesis. 

Actually, we find that $T$ can be negative, that is, the BS model may give
a better fit that the SB model. This is shown in the
tables after each figure, where we give the $T$-value 
for the example shown in the figure and the corresponding 
probability for acceptance. This is a priori surprising since,
by construction, the BS model is embedded in the SB model. The later
cannot in principle be worse that one of its special case.
In all cases, we find that the absolute value of $T$ is 
very large so as to give an acceptance of SB/BS of unity. 
The origins of these results are the following.
\begin{enumerate}
\item For numerical reasons, we cannot really
take the limit $\alpha \rightarrow \infty$ but only scan up to 
$\alpha_{max}=5$. Taking larger $\alpha$ values more often run 
into overflow problems (when the dominator in the SB fit is close to $0$).
The SB-fit can therefore never truely fit a BS model even in the deterministic 
case ($W=0$). Hence, the practical implementation of the SB model does
not fully encompass the BS model. As a consequence, it is natural that
the BS fit may be superior to the SB fit.

\item Due to the fluctuations in different realizations of $W(t)$, 
even exploring the limit $\alpha \rightarrow
\infty$ in the SB fits, randomness of $W$ still makes possible
that  $Q_{SB} >  Q_{BS}$.

\end{enumerate}
As a consequence, a true SB model can be
``mistaken'' for a BS model and vice versa for a true BS model. 

Consequently, we can not use the Likelihood-Ratio test to conclude reliably
on what is the correct model for a given time series. The above
procedure only tells us which one of the two models (BS versus SB) fits
better the data, given the implementation constraints. To obtain
a better sense of the statistical confidence with which one model is better
than the other one for a given time series, we need an additional test
that we know describe.

\subsection{Synthetic tests and bootstrapping \label{synthbook}}

In order to understand the meaning of the results presented below,
it is useful to test what our procedure described in the previous
section \ref{impl} gives for controlled cases.

As a starting point, let us take some typical parameters values 
$(\alpha=2.0, F=170, r=1.0 \times 10^{-4}, t_c=5000, \mu_0=5 \times 10^{-3}
, \sigma_0=1.0 \times 10^{-4})$ for the
Dow Jones Industrial Average (DJIA$_{1929}$) from 1/6 1927 to 30/9 1929, 
obtained by the
procedure of section \ref{impl}. Let us generate a synthetic SB time series from
equations 
(\ref{P},\ref{sb}) using one specific realisation of the random process $W(t)$. We call 
this synthetic generated SB bubble for $P_{\rm SBsyn}^1$.
We then consider $P_{\rm SBsyn}^1$ as an input data time 
series (called $P_{\rm data}$ in (\ref{Q})) and we want to see if our detection 
procedure can recognize this data series as a SB bubble. 
We therefore launch a
search for a time series $P(t)$ describing respectively a SB bubble
(\ref{P},\ref{sb})
and a BS bubble (\ref{P},\ref{bs}). Each search for a $P$  described in section
\ref{impl}
depends on the initial configuration of $W$, and we therefore launch 10
searches with different initial configurations of $W$.
For the case SB as well as
for BS we calculate $P^1_{\rm SB}$,
$P^1_{\rm BS}$ from the 10 searches with different inital configuration.
Another realisation $P_{\rm SBsyn}^2$ of the Dow bubble
is generated and again we launch 10 searches so as to calculate
$P^2_{\rm SB}$ and $P^2_{\rm BS}$. Making 10 realisations of the
Dow bubble, and defining $P_{\rm SB/BS} =  (1/10) \sum_{i=1}^{10} P^i_{\rm SB/BS}$
we finally get an estimate of $P_{\rm SB}=0.71$ and $P_{\rm BS}=0.28$.
As explained above, we
find always that one of the two models works overwhelmingly better,
corresponding to a large positive or negative $T$-value. 

This means that
the SB hypothesis is not necessary to describe 28\% of the realizations
generated in the SB regime and the BS description appears sufficient for them. 
In other words, a market phase
which is in a genuine SB regime by construction, 
characterized by strong nonlinear positive feedbacks according to the values
of its parameters (especially $1/\alpha$),
can be mistaken 28\% of the time as a ``normal'' BS regime!
Symmetrically we apply the same test in the BS regime
with parameters $(F=900, r=0.0002, \mu_0=0.001, \sigma_0=0.0001983)$ 
found for the Hang Seng
index from 1/6 1984 to 19/10 1987 before the crash in October 1987, we find that 
37\% of the BS replicas are detected as being in the BS regime
and 62\% are qualified as being in the SB regime. In contrast,
comparing the SB and BS regimes without stochasticity ($\sigma_0=0$), 
we find that a 100\% success rate in qualifying a SB as a SB and a BS as a BS.

Thus, it is clear that the failures to detect come from the stochastic character
of the realizations. And this should not be too surprising: indeed, a pure
BS random walk can give large spontaneous deviations that look like nonlinear
stochastic accelerations. Reciprocally, a SB realization may have its $W(t)$
wandering such that the denominator in the right-hand-side of (\ref{sb}) 
remains far from $0$ with $W(t)$ being only a relatively minor correction to the other terms
such that $B(t)$ has only a weak nonlinear correction compared with the standard
random walk. This is related to the general problem that there is no certainty
in assessing the random character of a finite string of numbers \cite{algocompl}:
for instance, the string head-head-head-head has exactly the same probability as 
head-tail-tail-head or any other for an unbiased coin toss. As the theory
of algorithm complexity has taught us, saying that the later
is more random that the former relies on the possibility to describe it with 
a shorter description or algorithm. Similarly, a SB is detected when it 
has a sufficiently strong linear structure that makes it unprobable to
result from a pure random realization. The unfortunate but unescapable
consequence is that the present stochastic rational expectation bubble
formulation makes the detection of a SB not certain, with
only the possibility of a probabilistic statement. 

Another natural test starts with the generation of many synthetic BS models
with fixed parameters obtained from the fit of the BS model to the data.
Then, fitting each of these synthetic BS replicas with the SB model,
we obtain a distribution of $1/\alpha$, which should be centered around $0$
and whose width should give us information on the uncertainty in recognizing
a BS bubble using a SB fit. We have found no significant difference with
this procedure, with a significant probability of taking a BS time series
for a SB model, as found with the other approach described above.

It is also interesting to note that, whenever we get
a predominant BS fit ($P_{BS} > P_{SB}$),
the range of $\alpha$ for SB is restricted to 
relatively small values. In contrast, for predominant SB fits, we get a broad
range of $\alpha$. Specifically, for bubbles clearly identified as BS, 
like e.g. HSI 1987, the range of $\alpha$ for the SB fits is ``small'' ($0.5 < 
\alpha < 2.5$ in that case), whereas for bubbles identified as
SB like e.g. DOW 1929 the range of $\alpha$ is ``large'' 
($0.5< \alpha < 5$ in that case). This again reflects the large
fluctuations inherent in the stochastic bubble model. 
The fact that $\alpha$ is smaller and clusters to smaller values for
predominant BS time series should give additional credit to the 
belief that those markets
are indeed BS-like since the well determined SB fits are inferior.
In contrast, the larger values of $\alpha$ 
for the predominant SB fits give credit to the 
belief that those markets are more SB-like.
The associated larger range of values $\alpha$ is a nuisance coming again
from the strongly stochastic character of the bubbles and the existence
of large deviations in the realizations of $W(t)$.

\section{Results of empirical tests on putative bubbles}

\begin{figure}
\begin{center}
\epsfig{file=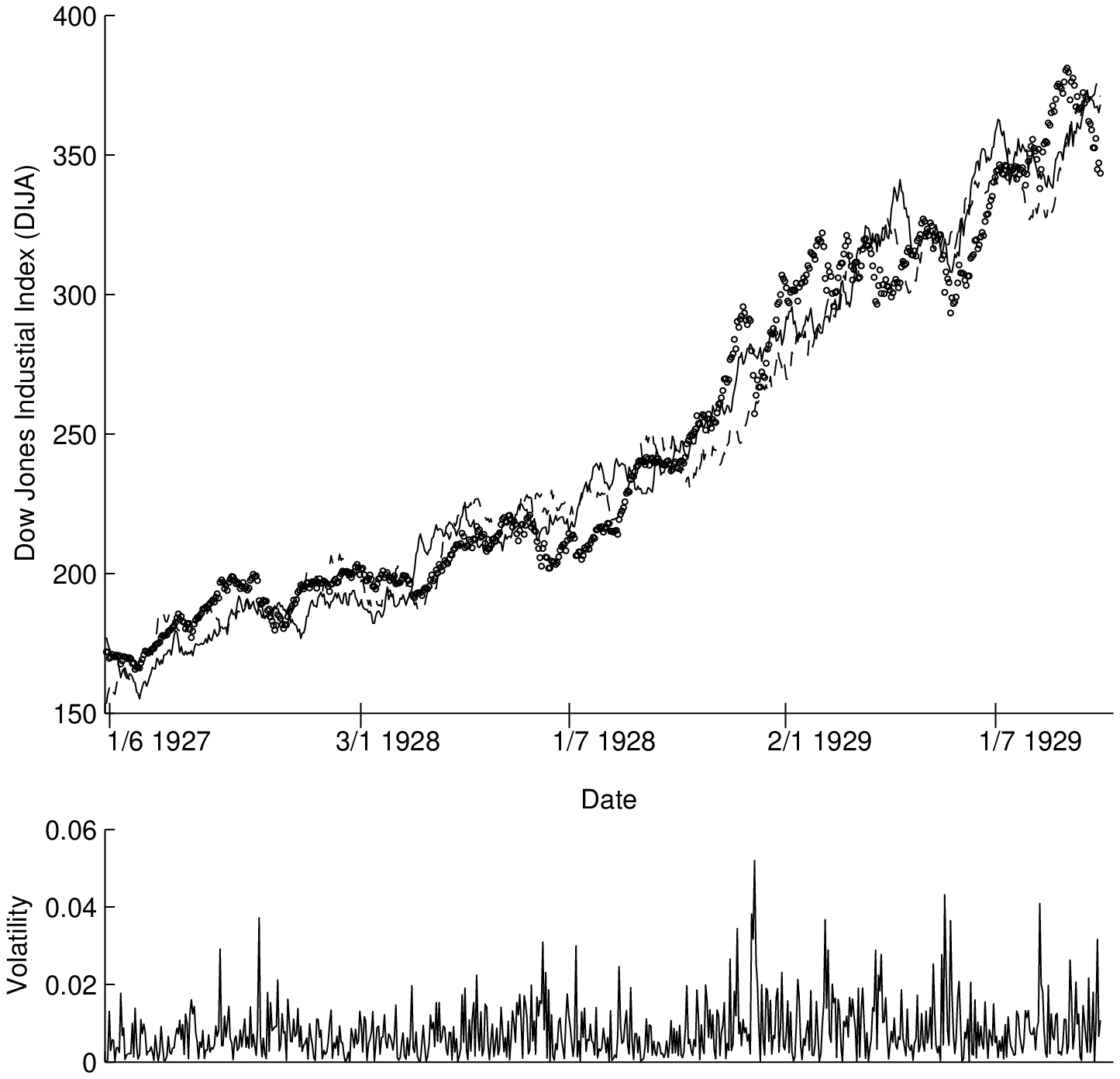,height=10cm,width=16cm}
\caption{\protect\label{fig_dow_1929} Upper panel:
Dow Jones Industrial Average (DJIA) Index $I(t)$ 
(open circles), together with 
the best fit with the SB (solid line) and BS (dashed line) models. 
Lower panel: Volatility defined as $|\ln[I(t)/I(t-1)]|$ as a function of time.
}
\end{center}
\end{figure}

Figure \ref{fig_dow_1929} shows the Dow Jones Industrial Average (DJIA) Index 
from 1/6 1927 to 1929 (open circles), a period preceding the 
famous crash of October 1929, together with 
the best fits with the SB (solid line) and BS (dashed line) models. 
The parameters of the fits are given in the table.
From the value of the $T$-statistics (\ref{T}), we obtained
a confidence level for the SB hypothesis of $p(T) = 1.0$.
Following the procedure described in section \ref{synthbook}, 
consisting in making 100 searches with the DJIA as the data 
time series
we find that
66\% of the searches are qualified as SB while 33\% are qualified as BS.

These two statistics confirm that the DJIA index from 1/6 1927 to 30/9 1929 exhibited
a pattern consistent with a ``fearful SB,'' to use the terminology
defined in section \ref{what}. The lower panel showing the
volatility of the DJIA index with time confirms that the increase of the price
was in fact accompanied by an overall increase in volatility.

\begin{figure}
\begin{center}
\epsfig{file=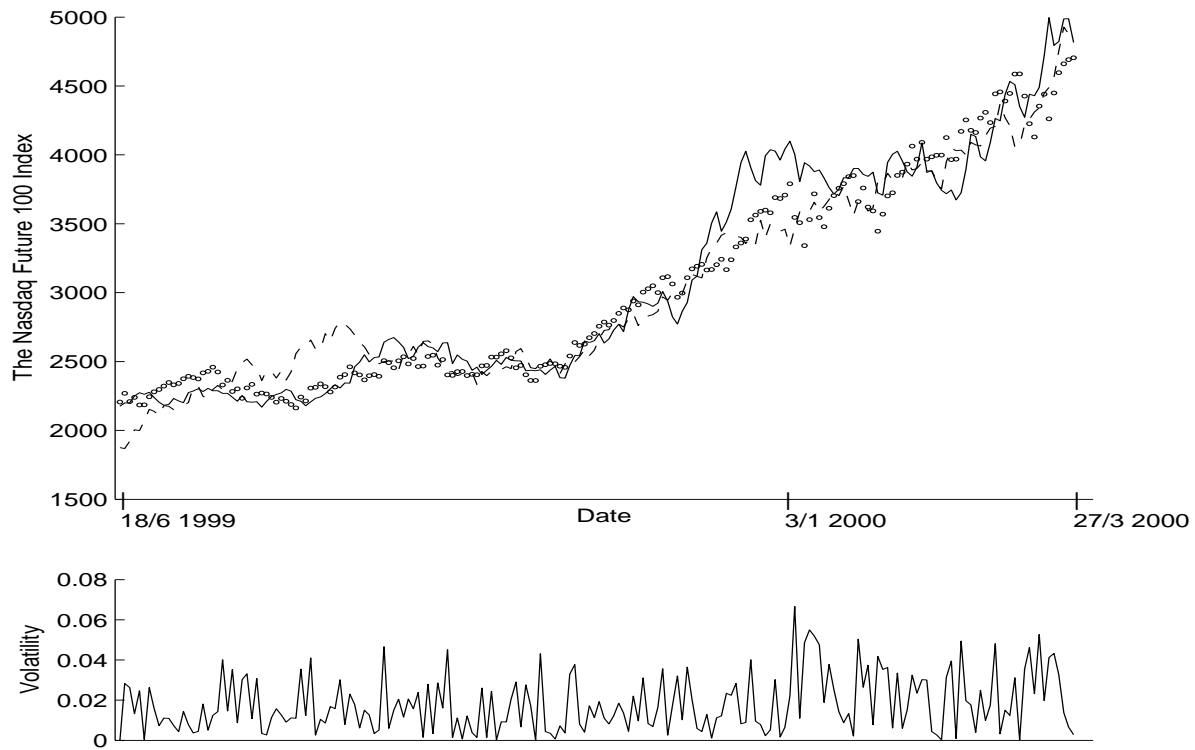,height=10cm,width=16cm}
\caption{\protect\label{fignas2000}  Same as Figure \ref{fig_dow_1929}
for the Nasdaq Future 100 Index from 18/6 1999 to 27/3 2000.
}
\end{center}
\end{figure}

Figure \ref{fignas2000} shows the Nasdaq Future 100 Index 
from 18/6 1999 to 27/3 2000 (open circles), a period often refered to as the new
economy information technology bubble, which preceded the crash in March-April 2000.
The best fits of one search with the SB (solid line) 
and BS (dashed line) models are also shown.
The parameters of the fits are given in the table.
From the value of the $T$-statistics (\ref{T}), we obtained
a confidence level for the SB hypothesis of $p(T) =0$.
As seen from the figure, the BS fit is clearly superior to the SB fit, 
especially at late times of the bubble, where the increasing volatility 
of the SB model does not match the approximately constant volatility observed
for the index. Furthermore, using the procedure described in section \ref{synthbook},
we find that 3\% of the searches are qualified as SB while a whooping
97\% are qualified as BS. The new economy information technology bubble cannot
be classified as a ``fearful SB.''

\begin{figure}
\begin{center}
\epsfig{file=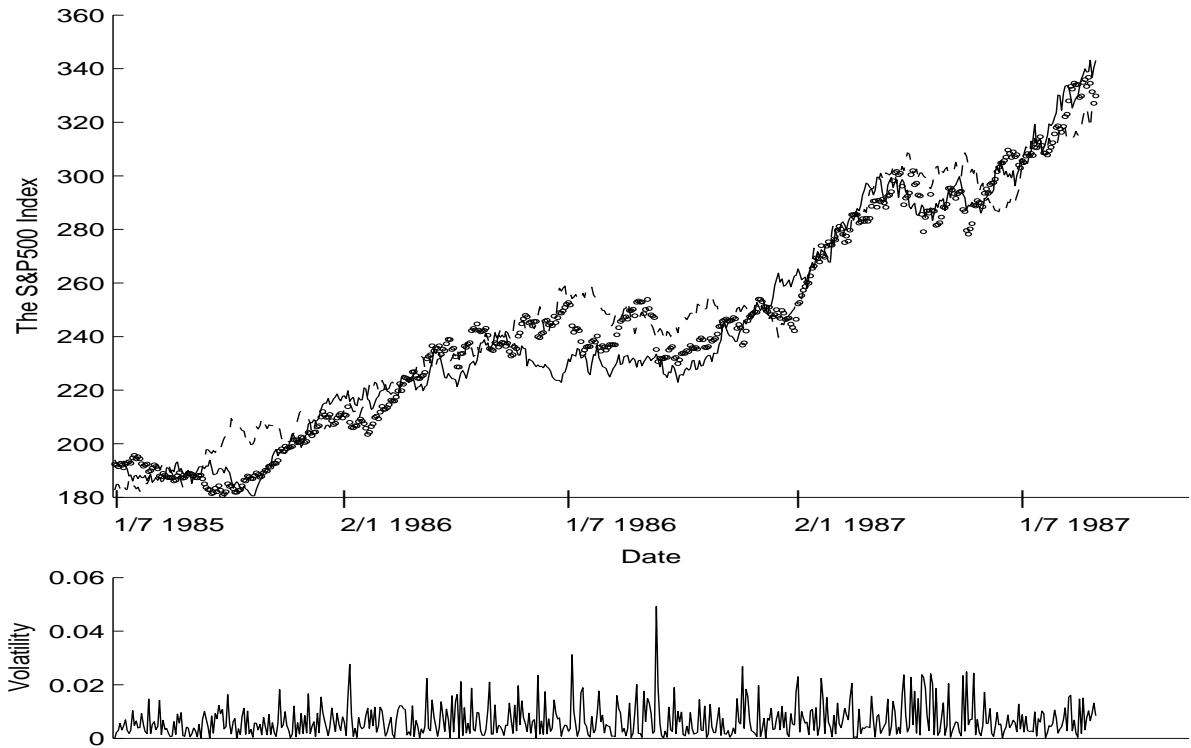,height=10cm,width=16cm}
\caption{\protect\label{fig_sp500_oct87} 
Same as Figure \ref{fig_dow_1929}
for the S\&P500 Index from 1/5 1987 to 31/8 1987.
}
\end{center}
\end{figure}

Figure \ref{fig_sp500_oct87}
shows the S\&P500 Index 
from 1/7 1985 to 31/8 1987 (open circles) before the crash 
on Black Monday, October 19, 1987, together with
the best fits with the SB (solid line) and BS (dashed line) models.
The parameters of the fits are given in the table.
From the value of the $T$-statistics (\ref{T}), we obtained
a confidence level for the SB hypothesis of $p(T) =1$.
Using the procedure described in section \ref{synthbook},
we find that 70\% of the synthetic bubbles are qualified as SB while
28\% are qualified as BS. The Black monday crash can thus be
classified as the end of a ``fearful SB.''

\begin{figure}
\begin{center}
\epsfig{file=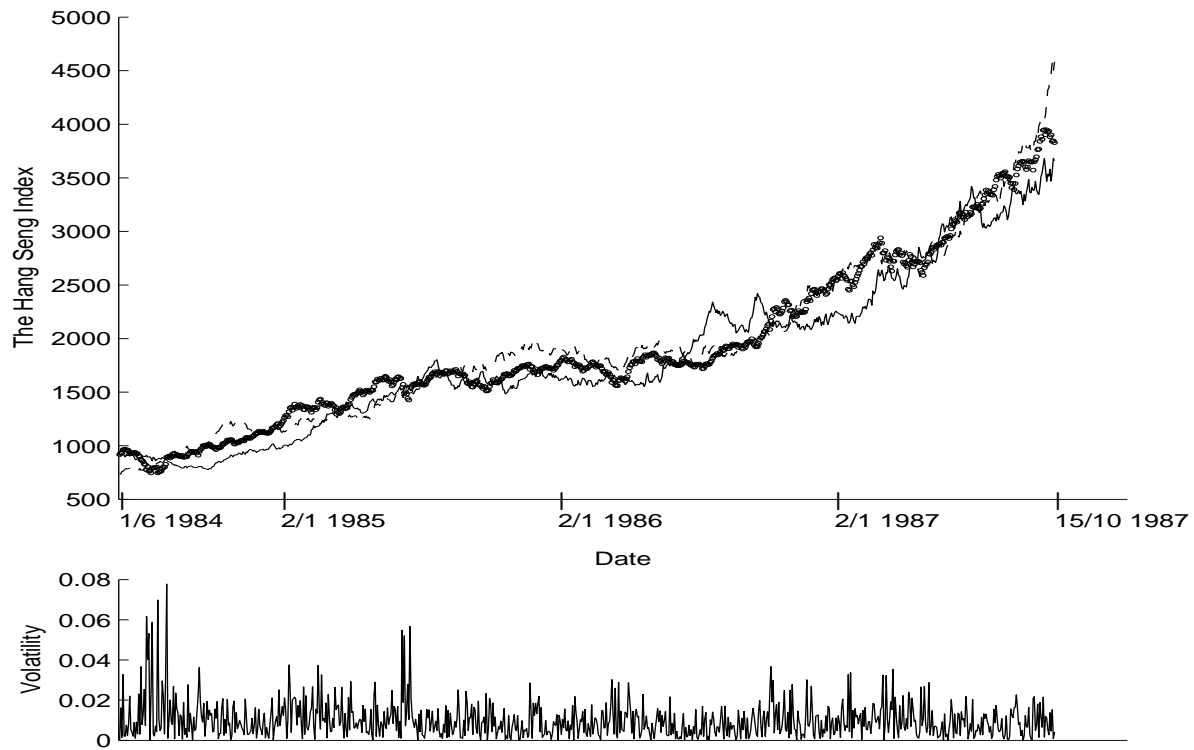,height=10cm,width=16cm}
\caption{\protect\label{fighsi_oct87} 
Same as Figure \ref{fig_dow_1929}
for the Hang Seng Index from 1/6 1984 to 19/10 1987.
}
\end{center}
\end{figure}

Figure \ref{fighsi_oct87} shows the Hong Kong Hang Seng Index 
from 1/6 1984 to 19/10 1987 (open circles) before its crash 
in October, 1987, together with
the best fits with the SB (solid line) and BS (dashed line) models.
The parameters of the fits are given in the table.
From the value of the $T$-statistics (\ref{T}), we obtained
a confidence level for the SB hypothesis of $p(T) =0$.
Using the procedure described in section \ref{synthbook},
we find that 0\% of the synthetic bubbles are qualified as SB while
100\% are qualified as BS. The accelerating price before this crash
cannot be classified as a ``fearful SB.''

\begin{figure}
\begin{center}
\epsfig{file=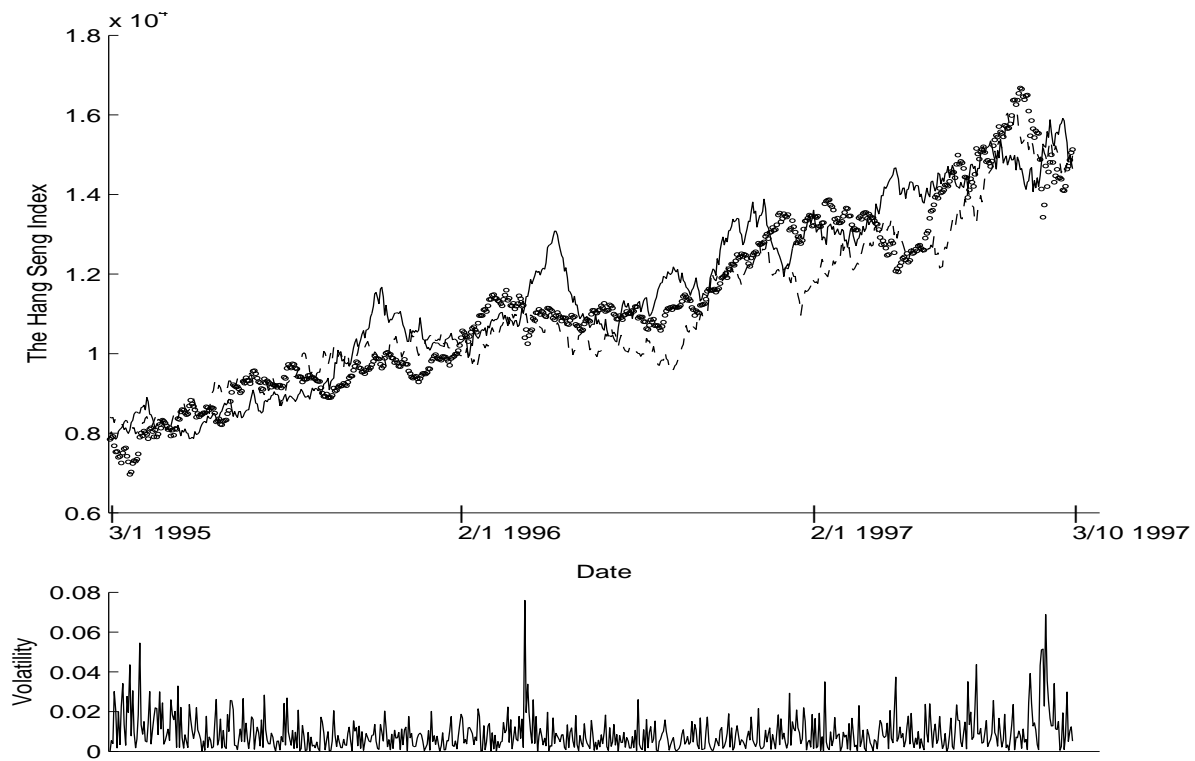,height=10cm,width=16cm}
\caption{\protect\label{fighsi_oct97} 
Same as Figure \ref{fig_dow_1929}
for the Hang Seng Index from 3/1 1995 to 3/10 1997.
}
\end{center}
\end{figure}

\begin{figure}
\begin{center}
\epsfig{file=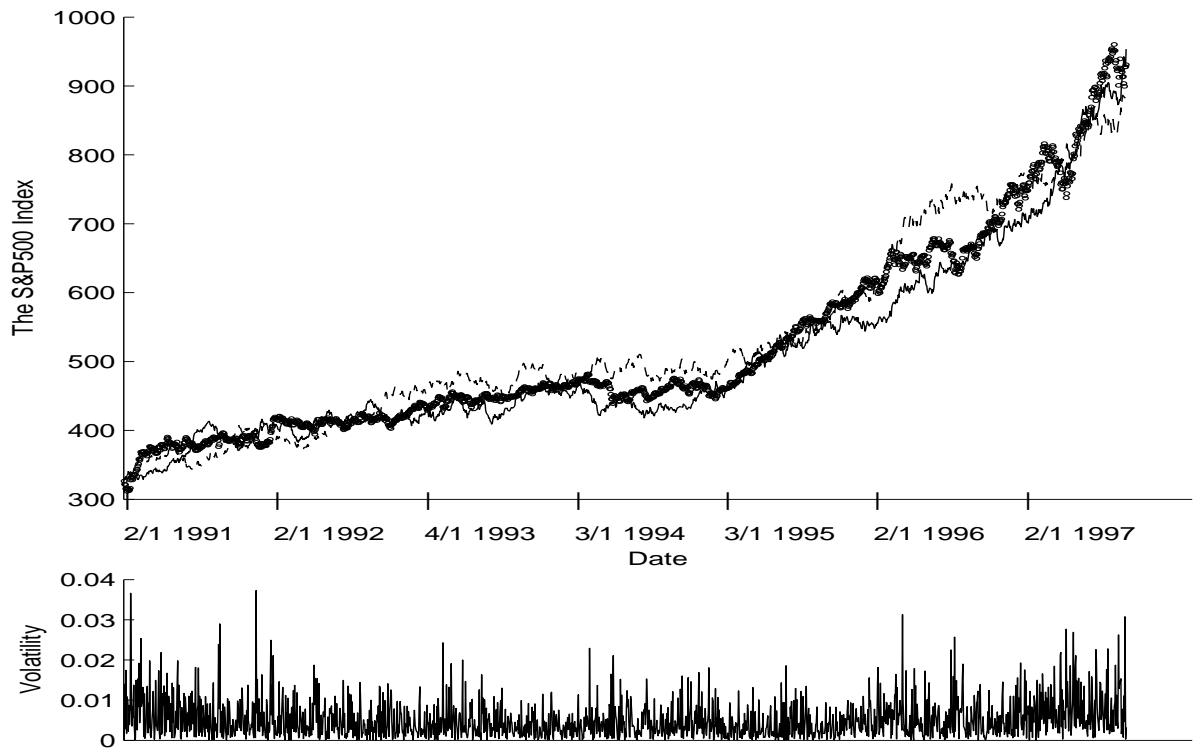,height=10cm,width=16cm}
\caption{\protect\label{figsp500_sep97} 
Same as Figure \ref{fig_dow_1929}
for the S\&P500 Index from 2/1 1991 to 4/9 1997.
}
\end{center}
\end{figure}

Figure \ref{fighsi_oct97} and 
figure \ref{figsp500_sep97} give the same information as for the
previous figures for the period from 3/1 1995 to 3/10 1997
preceding the crash of the Hang Seng index in October 1997
and for the period from 2/1 1991 to 4/9 1997 preceding the 
approximately $7\%$ one-day drop
of the S\&P500 index on October 27, 1997. 
From the value of the $T$-statistics (\ref{T}), we obtain
a confidence level for the SB hypothesis of $p(T) =0$ for the Hang Seng
index and of $p(T) =1$ for S\&P500 index.
Using the procedure described in section \ref{synthbook},
we find that 27\% of the searches are qualified as SB while
69\% are qualified as BS for the Hang Seng index, rejecting it
as a ``fearful SB.'' For the S\&P500 index,
we find that 75\% of the synthetic bubbles are qualified as SB while
25\% are qualified as BS, describing it as a ``fearful SB.''

\begin{figure}
\begin{center}
\epsfig{file=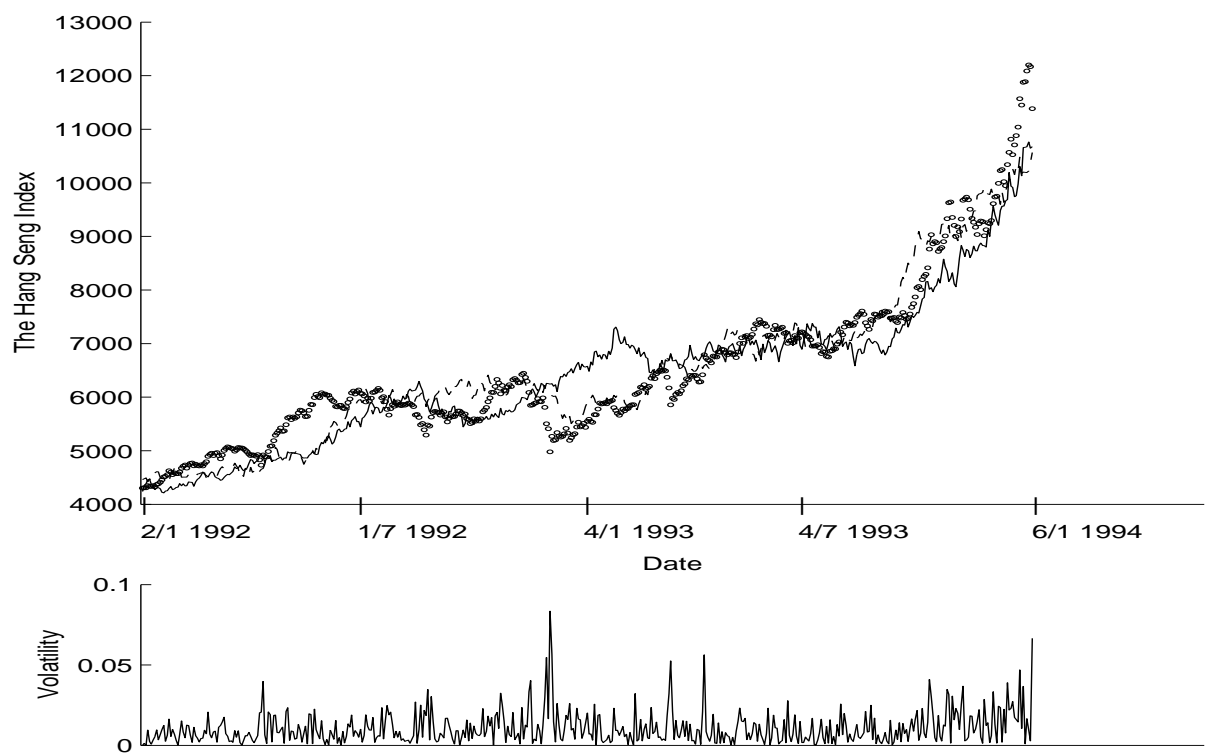,height=10cm,width=16cm}
\caption{\protect\label{fighsi_feb94} 
Same as Figure \ref{fig_dow_1929}
for the Hang Seng Index from 2/1 1992 to 6/1 1994.
}
\end{center}
\end{figure}

Figure \ref{fighsi_feb94} is the same for the Hang 
Seng index from 2/1 1992 to 6/1 1994, preceding the crash on February 1994. 
From the value of the $T$-statistics (\ref{T}), we obtained
a confidence level for the SB hypothesis of $p(T) =0$.
Using the procedure described in section \ref{synthbook},
we find that 27\% of the synthetic bubbles are qualified as SB while
73\% are qualified as BS for the Hang Seng index, rejecting it
as a ``fearful SB.'' 

\begin{figure}
\begin{center}
\epsfig{file=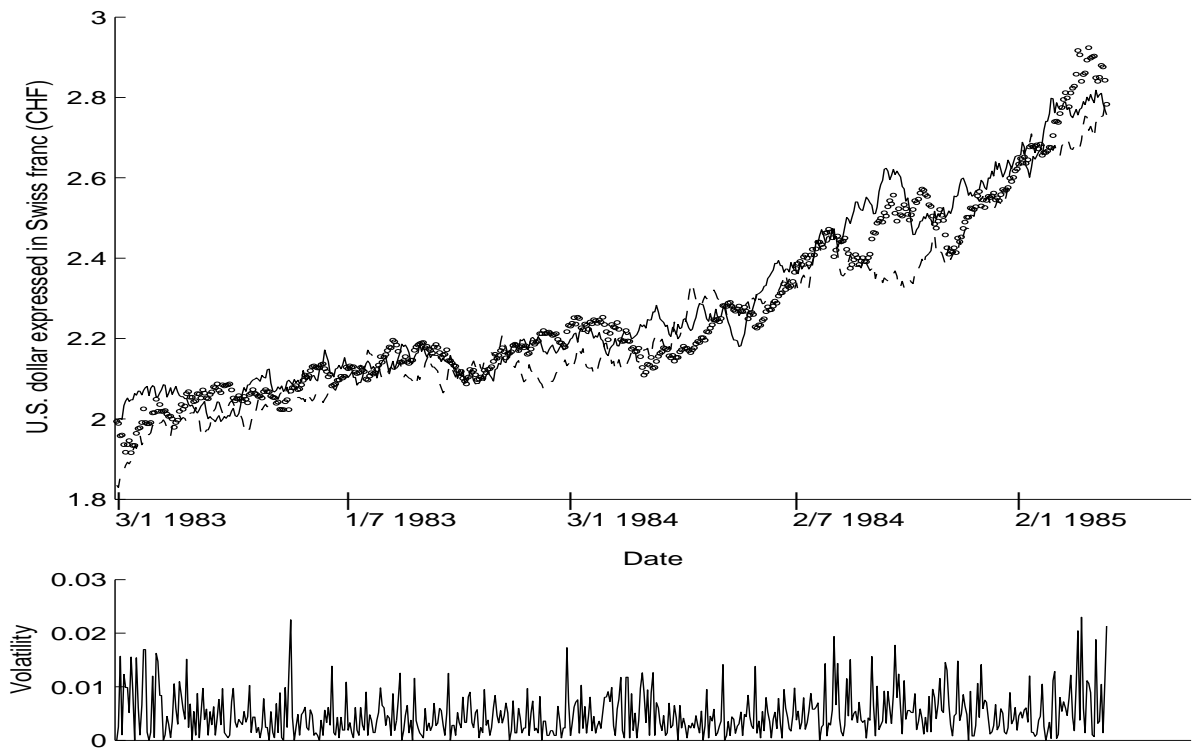,height=10cm,width=16cm}
\caption{\protect\label{figchf} 
Same as Figure \ref{fig_dow_1929}
for the US dollar in Swiss Francs (CHF) from 3/1 1983 to 19/3 1985.
}
\end{center}
\end{figure}

\begin{figure}
\begin{center}
\epsfig{file=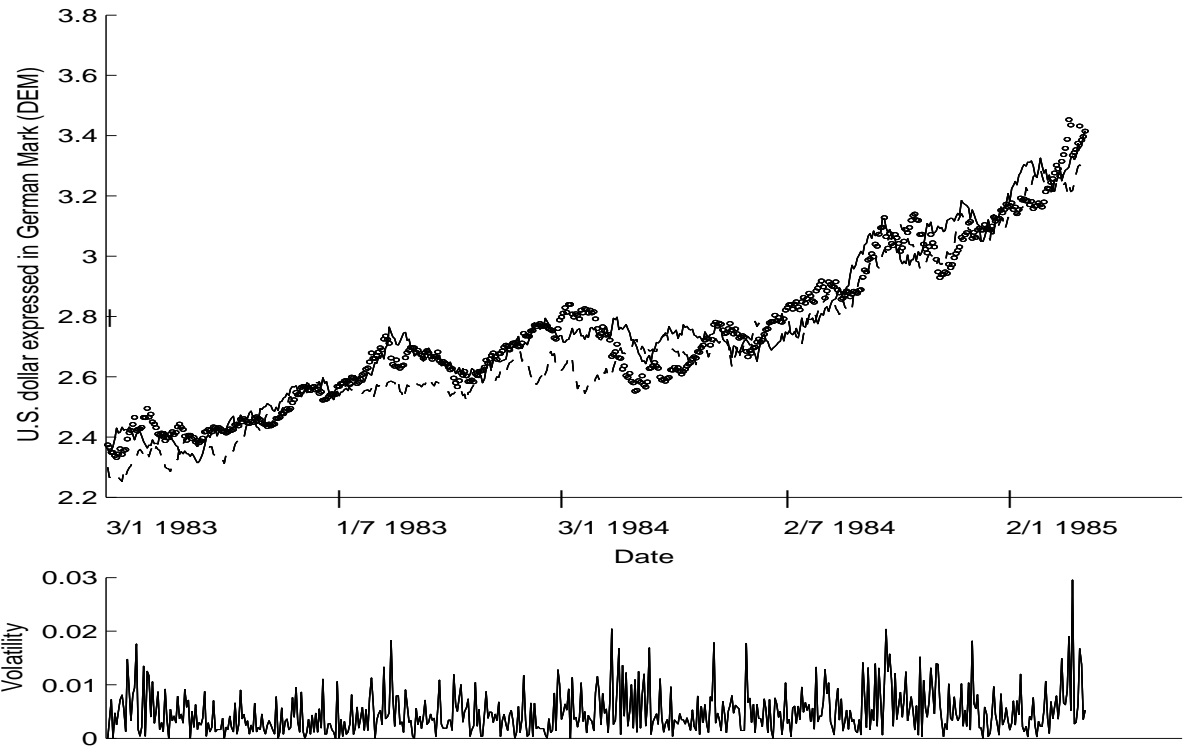,height=10cm,width=16cm}
\caption{\protect\label{figdem} Same as Figure \ref{fig_dow_1929}
for the US dollar in Deutch Marks (DEM) from 3/1 1983 to 8/3 1985.
}
\end{center}
\end{figure}

Figures \ref{figchf} and \ref{figdem} are the same for two
exchange rate time series preceding two crashes, namely the US dollar
expressed in Swiss Franc and in
German Mark, respectively, for the period from March 1983 to March 1985. 
From the value of the $T$-statistics (\ref{T}), we obtained
a confidence level for the SB hypothesis of $p(T) =1$ for US\$/CHF exchange rate
and of $p(T) =1$ for US\$/DEM echange rate.
Using the procedure described in section \ref{synthbook},
we find that 75\% of the searches are qualified as SB while
24\% are qualified as BS for the S\$/CHF exchange rate, describing it
as a ``fearful SB.'' For the US\$/DEM echange rate,
we find that 74\% of the synthetic bubbles are qualified as SB while
26\% are qualified as BS, describing it also as a ``fearful SB.''

\section{Synthesis and concluding remarks}

We have studied 9 time series which all ended in a crash or severe correction.
Many analysists have
classified these time series as bubbles, that is, exagerated
price increase over fundamental value (see for instance
\cite{bookcrash,SorJoh,SorPR} and references therein). We have examined these 9 events
through the lenses of our rational expectation bubble model.
The two fundamental ingredients of nonlinear positive feedback and of stochasticity
lead to the existence of two anomalous behaviors 
occurring simultaneously: (i) super-exponential price growth and (ii)
volatility growth. We refer to this behavior as the ``fearful singular bubble''
regime. Out of the 9 time series, we find that 5 pass our tests and
can be characterized as `fearful singular bubbles.'' The 4 other cases
which belong to a different class
are the information technology Nasdaq bubble and the three bubbles of the Hang Seng
index. Our model is not able to conclude whether or not these 4 cases are
bubbles. Many previous analyses have classified these 4 cases as bubbles,
but here we identify a distinguishing feature. According to our model, these
four bubbles (if we accept this classification) have developed with essentially
no significant increase of their volatility. They have been really
impressive price appreciation without detectable ``fear'' or ``worry''
as expressed by the volatility. In contrast, the other 5 bubbles
exhibited characteristics which make them recognizable in a framework
emphasizing the effect of positive feedback of prices acting both 
on returns and volatility. This means that bubble markets do not always 
``climb a wall of worry,'' as long as the ``worry'' is 
quantified by price volatilities. The information technology Nasdaq bubble and 
the three bubbles of the Hang Seng
index have been characterized by a remarkable ingenuous behavior on the part
of investors, apparently lacking any concern for possible associated risks
or any sense of fear. In other words, the
large price appreciations during the bubbles preceding the four crashes
were not coming with an increasing perception of risk on the part of 
investors. One could tentatively conclude that the market participants
were not mindful of the possibility of a crash risk looming. In contrast,
the five other bubbles are more in line with a rational expectation bubble
model, in which investors are at least partly aware of an increasing
looming risk.

{\bf Acknowledgements}: We thank Sam Salama for pointing out the references
of several of the quotes in the introduction.

\vskip 0.5cm

\pagebreak

\pagebreak
\begin{center}
\begin{tabular}{|c||c|c|c|c|c|c|}
\hline
\multicolumn{7}{|c|}{Dow Jones Industrial Average from 1/6 1927 to 30/9 1929}
\\
\hline
\multicolumn{7}{|c|}{Percentage of 100 searches $P_{SB}=0.66$, $P_{BS}=0.33$}
\\ \hline
\hline
Parameters of the curves in the figure & $F$ & $r$ & $\sigma$ & $\mu$ & $\alpha$ & $T_c$
\\ \hline
SB & 167.93 & 4.0 $\times 10^{-5}$ & 6.49 $\times 10^{-5}$ & 
7.65 $\times 10^{-4}$ & 1.5 & 1929
\\ \hline
BS & 167.47 & 3.2 $\times 10^{-4}$ & 1.09 $\times 10^{-4}$ & 
6.82 $\times 10^{-3}$ & - & - 
\\ \hline
\multicolumn{7}{|c|}{T=119.42 \ \ Prob.(SB) $>$  0.9999}
\\ \hline
\end{tabular}
\end{center}

\pagebreak
\begin{center}
\begin{tabular}{|c||c|c|c|c|c|c|}
\hline
\multicolumn{7}{|c|}{Nasdaq Future 100 Index from 18/6 1999 to 27/3 2000}
\\
\hline
\multicolumn{7}{|c|}{Percentage of 100 searches $P_{SB}=0.03$, $P_{BS}=0.97$}
\\ \hline
\hline
Parameters of the curves in the figure & $F$ & $r$ & $\sigma$ & $\mu$ & $\alpha$ & $T_c$
\\ \hline
SB & 1262.08 & 2.0 $\times 10^{-5}$ & 8.89 $\times 10^{-5}$ & 
1.55 $\times 10^{-3}$ & 1.0 & 368.34
\\ \hline
BS & 1845.20 & 1.60 $\times 10^{-4}$ & 5.10 $\times 10^{-4}$ & 
3.73 $\times 10^{-3}$ & - & - 
\\ \hline
\multicolumn{7}{|c|}{T=-45.30 \ \ Prob.(BS) $>$  0.9999}
\\ \hline
\end{tabular}
\end{center}

\pagebreak
\begin{center}
\begin{tabular}{|c||c|c|c|c|c|c|}
\hline
\multicolumn{7}{|c|}{S\&P 500 Index from 1/7 1985 to 31/8 1987}
\\
\hline
\multicolumn{7}{|c|}{Percentage of 100 searches $P_{SB}=0.70$, $P_{BS}=0.28$}
\\ \hline
\hline
Parameters of the curves in the figure & $F$ & $r$ & $\sigma$ & $\mu$ & $\alpha$ & $T_c$
\\ \hline
SB & 110.64 & 2.0 $\times 10^{-5}$ & 4.59 $\times 10^{-5}$ & 
7.36 $\times 10^{-4}$ & 3.0 & 3390.54
\\ \hline
BS & 185.06 & 2.0 $\times 10^{-4}$ & 7.81 $\times 10^{-5}$ & 
9.65 $\times 10^{-4}$ & - & - 
\\ \hline
\multicolumn{7}{|c|}{T=183.71 \ \ Prob.(SB) $>$  0.9999}
\\ \hline
\end{tabular}
\end{center}

\pagebreak
\begin{center}
\begin{tabular}{|c||c|c|c|c|c|c|}
\hline
\multicolumn{7}{|c|}{Hang Seng Index from 1/6 1984 to 19/10 1987}
\\
\hline
\multicolumn{7}{|c|}{Percentage of 100 searches $P_{SB}=0.01$, $P_{BS}=0.99$}
\\ \hline
\hline
Parameters of the curves in the figure & $F$ & $r$ & $\sigma$ & $\mu$ & $\alpha$ & $T_c$
\\ \hline
SB & 448.53 & 3.2 $\times 10^{-4}$ & 8.17 $\times 10^{-5}$ & 
7.59 $\times 10^{-4}$ & 2.5 & 2473.03
\\ \hline
BS & 732.50 & 6.4 $\times 10^{-4}$ & 1.98 $\times 10^{-4}$ & 
9.14 $\times 10^{-4}$ & - & - 
\\ \hline
\multicolumn{7}{|c|}{T=-535.38 \ \ Prob.(BS) $>$  0.9999}
\\ \hline
\end{tabular}
\end{center}

\pagebreak
\begin{center}
\begin{tabular}{|c||c|c|c|c|c|c|}
\hline
\multicolumn{7}{|c|}{Hang Seng Index from 3/1 1995 to 3/10 1997}
\\
\hline
\multicolumn{7}{|c|}{Percentage of 100 searches $P_{SB}=0.27$, $P_{BS}=0.69$}
\\ \hline
\hline
Parameters of the curves in the figure & $F$ & $r$ & $\sigma$ & $\mu$ & $\alpha$ & $T_c$
\\ \hline
SB & 7075.35 & 8.0 $\times 10^{-5}$ & 9.50 $\times 10^{-5}$ & 
6.81 $\times 10^{-4}$ & 1.5 & 2055.85 
\\ \hline
BS & 8637.04 & 1.6 $\times 10^{-4}$ & 1.60 $\times 10^{-4}$ & 
8.06 $\times 10^{-3}$ & - & - 
\\ \hline
\multicolumn{7}{|c|}{T=-151.21 \ \ Prob.(BS) $>$  0.9999}
\\ \hline
\end{tabular}
\end{center}

\pagebreak
\begin{center}
\begin{tabular}{|c||c|c|c|c|c|c|}
\hline
\multicolumn{7}{|c|}{S\&P 500 Index from 2/1 1991 to 4/9 1997}
\\
\hline
\multicolumn{7}{|c|}{Percentage of 100 searches $P_{SB}=0.75$, $P_{BS}=0.25$}
\\ \hline
\hline
Parameters of the curves in the figure & $F$ & $r$ & $\sigma$ & $\mu$ & $\alpha$ & $T_c$
\\ \hline
SB & 436.18 & 3.2 $\times 10^{-4}$ & 5.18 $\times 10^{-5}$ & 
3.03 $\times 10^{-4}$ & 5 & 17481.7
\\ \hline
BS & 329.97 & 3.2 $\times 10^{-4}$ & 4.98 $\times 10^{-4}$ & 
3.01 $\times 10^{-4}$ & - & - 
\\ \hline
\multicolumn{7}{|c|}{T=696.78 \ \ Prob.(SB) $>$  0.9999}
\\ \hline
\end{tabular}
\end{center}

\pagebreak
\begin{center}
\begin{tabular}{|c||c|c|c|c|c|c|}
\hline
\multicolumn{7}{|c|}{Hang Seng Index from 2/1 1992 to 6/1 1994}
\\
\hline
\multicolumn{7}{|c|}{Percentage of 100 searches $P_{SB}=0.26$, $P_{BS}=0.74$}
\\ \hline
\hline
Parameters of the curves in the figure & $F$ & $r$ & $\sigma$ & $\mu$ & $\alpha$ & $T_c$
\\ \hline
SB & 3454.43 & 3.2 $\times 10^{-4}$ & 7.30 $\times 10^{-5}$ & 
8.90 $\times 10^{-4}$ & 1.0 & 902.14 
\\ \hline
BS & 4507.98 & 3.2 $\times 10^{-4}$ & 2.12 $\times 10^{-4}$ & 
1.62 $\times 10^{-3}$ & - & - 
\\ \hline
\multicolumn{7}{|c|}{T=-283.28 \ \ Prob.(BS) $>$  0.9999}
\\ \hline
\end{tabular}
\end{center}

\pagebreak
\begin{center}
\begin{tabular}{|c||c|c|c|c|c|c|}
\hline
\multicolumn{7}{|c|}{Swiss Francs (CHF) from 3/1 1983 to 19/3 1985}
\\
\hline
\multicolumn{7}{|c|}{Percentage of 100 searches $P_{SB}=0.75$, $P_{BS}=0.24$}
\\ \hline
\hline
Parameters of the curves in the figure & $F$ & $r$ & $\sigma$ & $\mu$ & $\alpha$ & $T_c$
\\ \hline
SB & 2.63 & 4.0 $\times 10^{-5}$ & 4.37 $\times 10^{-5}$ & 
5.84 $\times 10^{-4}$ & 3.5 & 6486.15 
\\ \hline
BS & 1.89 & 3.1.6 $\times 10^{-3}$ & 4.02 $\times 10^{-5}$ & 
4.44 $\times 10^{-4}$ & - & - 
\\ \hline
\multicolumn{7}{|c|}{T=43.02 \ \ Prob.(SB) $>$  0.9999}
\\ \hline
\end{tabular}
\end{center}

\pagebreak
\begin{center}
\begin{tabular}{|c||c|c|c|c|c|c|}
\hline
\multicolumn{7}{|c|}{Deutch Marks (DEM) from 3/1 1983 to 8/3 1985}
\\
\hline
\multicolumn{7}{|c|}{Percentage of 100 searches $P_{SB}=0.74$, $P_{BS}=0.26$}
\\ \hline
\hline
Parameters of the curves in the figure & $F$ & $r$ & $\sigma$ & $\mu$ & $\alpha$ & $T_c$
\\ \hline
SB & 2.17 & 3.2 $\times 10^{-4}$ & 3.37 $\times 10^{-5}$ & 
3.16 $\times 10^{-4}$ & 2.0 & 6036.89
\\ \hline
BS & 2.29 & 1.6 $\times 10^{-4}$ & 3.95 $\times 10^{-5}$ & 
5.07 $\times 10^{-3}$ & - & - 
\\ \hline
\multicolumn{7}{|c|}{T=256.898987 \ \ Prob.(SB) $>$  0.9999}
\\ \hline
\end{tabular}
\end{center}

\end{document}